\documentclass[acmsmall,sigconf]{acmart}
\usepackage{comment}
\usepackage{float}

\AtBeginDocument{%
  \providecommand\BibTeX{{%
    \normalfont B\kern-0.5em{\scshape i\kern-0.25em b}\kern-0.8em\TeX}}}

\copyrightyear{2021}
\acmYear{2021}
\setcopyright{acmlicensed}
\acmConference[CHI '21]{CHI Conference on Human Factors in Computing Systems}{May 8--13, 2021}{Yokohama, Japan}
\acmBooktitle{CHI Conference on Human Factors in Computing Systems (CHI '21), May 8--13, 2021, Yokohama, Japan}
\acmPrice{15.00}
\acmDOI{10.1145/3411764.3445170}
\acmISBN{978-1-4503-8096-6/21/05}

\begin{document}

\title{Extended Reality (XR) Remote Research: a Survey of Drawbacks and Opportunities}


\author{Jack Ratcliffe}
\authornote{Both authors contributed equally to this research.}
\orcid{0000-0001-5729-3252}
\affiliation{%
  \institution{Queen Mary, University of London}
  \streetaddress{Mile End Road}
  \city{London}
      \country{UK}
  \postcode{E1 4NS}
}

\author{Francesco Soave}
\authornotemark[1]
\orcid{0000-0002-7136-2135}
\affiliation{%
  \institution{Queen Mary, University of London}
  \streetaddress{Mile End Road}
  \city{London}
      \country{UK}
  \postcode{E1 4NS}
}

\author{Nick Bryan-Kinns}
\affiliation{%
  \institution{Queen Mary University of London}
  \streetaddress{Mile End Road}
  \city{London}
      \country{UK}
  \postcode{E1 4NS}
}

\author{Laurissa Tokarchuk}
\affiliation{%
  \institution{Queen Mary University of London}
  \streetaddress{Mile End Road}
  \city{London}
    \country{UK}
  \postcode{E1 4NS}
}

\author{Ildar Farkhatdinov}
\affiliation{%
  \institution{Queen Mary University of London}
  \streetaddress{Mile End Road}
  \city{London}
      \country{UK}
  \postcode{E1 4NS}
}




\begin{abstract}
  Extended Reality (XR) technology - such as virtual and augmented reality - is now widely used in Human Computer Interaction (HCI), social science and psychology experimentation. However, these experiments are predominantly deployed in-lab with a co-present researcher. Remote experiments, without co-present researchers, have not flourished, despite the success of remote approaches for non-XR investigations. This paper summarises findings from a 30-item survey of 46 XR researchers to understand perceived limitations and benefits of remote XR experimentation. Our thematic analysis identifies concerns common with non-XR remote research, such as participant recruitment, as well as XR-specific issues, including safety and hardware variability. We identify potential positive affordances of XR technology, including leveraging data collection functionalities builtin to HMDs (e.g. hand, gaze tracking) and the portability and reproducibility of an experimental setting. We suggest that XR technology could be conceptualised as an interactive technology and a capable data-collection device suited for remote experimentation.
 
\end{abstract}



\ccsdesc[500]{Human-centered computing~Mixed / augmented reality}
\ccsdesc[500]{Human-centered computing~Virtual reality}

\keywords{Extended Reality, Virtual Reality, Augmented Reality, literature review, expert interviews}

\maketitle

\section{Introduction}

Extended reality (XR) technology  - such as virtual, augmented, and mixed reality - is increasingly being examined and utilised by researchers in the HCI and other research communities due to its potential for creative, social and psychological experiments~\cite{blascovich2002immersive}. Many of these studies take place in laboratories with the co-presence of the researcher and the participant~\cite{kourtesis2020guidelines}. The XR research community has been slow to embrace recruiting remote participants to take part in studies running outside of laboratories - a technique which has proven useful for non-XR HCI, social and psychological research~\cite{preece2016citizen}\cite{paolacci2010running}. However, the current Covid-19 pandemic has highlighted the importance and perhaps necessity of understanding and deploying remote recruitment methods within XR research.

There is also limited literature \textit{about} remote XR research, although what reports exist suggest that the approach shows promise: data-collection is viable~\cite{7383331}, results are similar to those found in-lab~\cite{mottelson2017virtual} even when the participants are unsupervised~\cite{huber2020conducting}, and recruiting is possible \cite{ma2018web}. Researchers have also suggested using existing communities for these technologies, such as customisable social VR experiences, as combined platforms for recruitment and experimentation \cite{saffo2020crowdsourcing}. With the increasingly availability of consumer XR devices (estimates show five million high-end XR HMDs sold in 2020, raising to 43.5 million by 2025\cite{tankovska_2020}), and health and safety concerns around in-lab experimentation, particularly for research involving head-mounted displays (HMDs), it seems an important time to understand the conceptions around remote research from researchers who use XR technologies. 

This paper outlines the methodology and results from the first (that we are aware of) survey of XR researchers regarding remote XR research. The results have been derived from 46 respondents answering 30 questions regarding existing research practice. It offers three core contributions: (1) we summarise existing research on conducting remote XR experiments. (2) We provide an overview of the status quo, showing that many of the concerns regarding remote XR are those also applicable to other remote studies; and that the unique aspects of remote XR research could offer more benefits than drawbacks. (3) We set out recommendations for advancing remote XR research, and outline important questions that should be answered to create an evidence-backed experimentation process.

\section{Literature}
We present a literature review of relevant publications on XR research, remote research and remote XR research. We use "XR" as the umbrella term for virtual reality (VR), augmented reality (AR) and mixed reality (MR)~\cite{ludlow2015}. This space is also sometimes referred to as spatial or immersive computing. 

The chapter is organised in three parts. First, we explore conventional XR experiments under `normal' conditions (e.g. in laboratory and\/or directly supervised by the researcher). We then summarise existing literature on remote experiments in XR research. Finally, we report the main findings in previous publications on remote data collection and experimentation.

\subsection{Conventional XR experiments}

\subsubsection{Experiment types and fields of interest}

According to Suh and Prohpet's 2018 systematic literature review~\cite{suh2018state}, XR experiments involving human participants can broadly be categorised into two groups: (1) studies about XR, and (2) studies about \emph{using} XR. The first group focuses on the effects of XR system features on the user experience (e.g. if enhancing embodiment could affect presence outcomes~\cite{ratcliffe}), whereas the second category examines how the use of an XR technology modifies a measurable user attribute (e.g. if leveraging XR embodiment could affect learning outcomes~\cite{ratcliffeanalysing}).
Across these categories there have been a variety of explorations on different subjects and from different academic fields. These include social psychological~\cite{blascovich2002immersive}, including social facilitation–inhibition~\cite{hoyt2003social}, conformity and social comparison~\cite{blascovich2002social}, social identity~\cite{kilteni2013drumming}; neuroscience and neuropsychology~\cite{kourtesis2020guidelines}, visual perception~\cite{wilson2015use}, multisensory integration~\cite{choi2016multisensory}, proxemics~\cite{sanz2015virtual}, spatial cognition~\cite{waller2007hive}, education and training~\cite{radianti2020systematic}, therapeutic applications~\cite{freeman2018automated}, pain remediation~\cite{gromala2015virtual},  motor control~\cite{connelly2010pneumatic}, terror management~\cite{josman2006busworld} and media effects such as presence~\cite{bailey2012presence}. 

The theoretical approaches behind these studies are also disparate, including theories such as conceptual blending, cognitive load, constructive learning, experiential learning, flow, media richness, motivation, presence, situated cognition, the stimuli-organism-response framework and the technology acceptance model~\cite{suh2018state}.

\subsubsection{Data collection, approaches and techniques}

According to Suh and Prophet's meta-analysis, the majority of XR research explorations have been experiments (69\%)~\cite{suh2018state}. Other types of explorations include surveys (24\%), interviews (15\%) and case studies (9\%). These approaches have been used both alone and in combination with each other. Data collection methods are predominantly quantitative (78\%), although qualitative and mixed approaches are also used. Another systematic review of XR research (focused on higher education)~\cite{radianti2020systematic} adds focus group discussion and observation as research methods, and presents two potential subcategories for experiments: mobile sensing and "interaction log in VR app", in which the XR application logs the user's activities and the researcher uses the resulting log for analysis.

The types of data logging found in XR experiments are much the same as those listed in Weibel's exploration of physiological measures in non-immersive virtual reality~\cite{weibel2018virtual}, with studies using skin conductance~\cite{yuan2010rubber}, heart rate~\cite{egan2016evaluation}, blood pressure~\cite{hoffman2003immersive}, as well as electroencephalogram (EEG)~\cite{amores2018promoting}. Built-in inertial sensors that are integral to providing an XR experience, such as head and hand position for VR HMDs, have also been widely used for investigations, including posture assessment~\cite{brookes2019studying}, head interaction tracking~\cite{zhang2018human}, gaze and loci of attention~\cite{piumsomboon2017exploring} and gesture recognition~\cite{kehl2004real}, while velocity change~\cite{warriar2019modelling} has also been used in both VR and AR interventions.

\subsubsection{Benefits of XR experiments}

There are many suggested benefits to using XR technology as a research tool: it allows researchers to \textit{control the mundane-realism trade-off}~\cite{aronson1969theory} and thus increase the extent to which an experiment is similar to situations encountered in everyday life without sacrificing experimental control~\cite{blascovich2002immersive}; to create \textit{powerful sensory illusions within a  controlled environment (particularly in VR)}, such as illusions of self-motion and influence the proprioceptive sense~\cite{soave2020vection}; \textit{improve replication}~\cite{blascovich2002immersive} by making it easier to recreate entire experimental environments; and allow \textit{representative samples}\cite{blascovich2002immersive} to experience otherwise inaccessible environments, when paired with useful distribution and recruitment networks. 

\subsubsection{Challenges of XR experiments}

Pan~\cite{pan2018and} explored some of the challenges facing experiments in virtual worlds, which continue to be relevant in immersive XR explorations. These include the challenge of ensuring the \textit{experimental design} is relevant for each technology and subject area; ensuring a consistent feeling of \textit{self‐embodiment} to ensure engaged performance~\cite{kilteni2012sense}; avoid \textit{uncanny valley}, in which characters which look nearly-but-not-quite human are judged as uncanny and are aversive for participants~\cite{mori2012uncanny}; \textit{simulation sickness} and nausea during VR experiences\cite{moss2011characteristics}; \textit{cognitive load}~\cite{sweller2010cognitive} which may harm participation results through over-stimulation, particularly in VR~\cite{steed2016impact}~\cite{makransky2019adding}; \textit{novelty effects} of new technology interfering with results~\cite{clark1992research}~\cite{ely1994educational}; and \textit{ethics}, especially where experiences in VR could lead to changes in participants’ behaviour and attitude in their real life~\cite{banakou2016virtual} and create false memories~\cite{segovia2009virtually}.

\subsection{Remote XR experiments}

There has been little research into remote XR experimentation, particularly for VR and AR HMDs. By remote, we mean any experiment that takes place outside of a researcher-controlled setting. This is distinct from field or in-the-wild research, which is research "that seeks to understand new technology interventions in everyday living"~\cite{rogers2017research}, and so is dependent on user context. These definitions are somewhat challenged in the context of remote VR research, as for VR, remote and field/in-the-wild are often the same setting, as the location where VR is most used outside the lab is also where it is typically experienced (e.g. home users, playing at home~\cite{ma2018web}). For AR, there is a greater distinction between remote, which refer to any AR outside of the controlled setting of the lab; and field/in-the-wild, which require a contextual deployment.

In terms of remote XR research outcomes, Mottelson and Hornbæk~\cite{mottelson2017virtual} directly compared in-lab and remote VR experiment results. They found that while the differences in performance between the in-lab and remote study were substantial, there were no significant differences between effects of experimental conditions. Similarly, Huber and Gajos explored uncompensated and unsupervised remote VR samples and were able to replicate key results from the original in-lab studies, although with smaller effect sizes~\cite{huber2020conducting}. Finally, Steed et al. showed that collecting data in the wild is feasible for virtual reality systems~\cite{7383331}.

Ma et al.~\cite{ma2018web} is perhaps the first published research on recruiting remote participants for VR research. The study, published in 2018, used the Amazon Mechanical Turk (AMT) crowdsourcing platform, and received 439 submissions over a 13-day period, of which 242 were eligible. The participant demographics did not differ significantly from previously reported demographics of AMT populations in terms of age, gender, and household income. The notable difference was that the VR research had a higher percentage of U.S.-based workers compared to others. The study also provides insight into how remote XR studies take place: 98\% of participants took part at home, in living rooms (24\%), bedrooms (18\%), and home offices (18\%). Participants were typically alone (84\%) or in the presence of one (14\%) or two other people (2\%). Participants reported having “enough space to walk around” (81\%) or “run around (10\%)”. Only 6\% reported that their physical space would limit their movement.

While Ma et al's work is promising in terms of reaching a representative sample and the environment in which participants take part in experiments, it suggests a difficulty in recruiting participants with high-end VR systems, which allow six-degrees of freedom (the ability to track user movement in real space) and leverage embodied controllers (e.g. Oculus Rift, HTC Vice). Only 18 (7\%) of eligible responses had a high-end VR system. A similar paucity of high-end VR equipment was found by Mottelson and Hornbæk~\cite{mottelson2017virtual}, in which 1.4\% of crowdworkers had access to these devices (compared to 4.5\% for low-end devices, and 83.4\% for Android smartphones). This problem is compounded if we consider Steed et al's finding that only 15\% of participants provide completed sets of data~\cite{7383331}.

An alternative approach to recruiting participants is to create experiments inside existing communities of XR users, such as inside the widely-used VR Chat software~\cite{saffo2020crowdsourcing}. This allows researchers to enter into existing communities of active users, rather than attempt to establish their own. However, there are significant limitations for building experiments on platforms not designed for experimentation, such as programming limitations, the ability to communicate with outside services for data storage, and the absence of bespoke hardware interfaces. 

\subsection{Remote data collection and experimentation}

\subsubsection{Validity, benefits, drawbacks and differences}

Using networks for remote data collection from human participants has been proven valid in some case studies~\cite{gosling2004should,krantz2000validity}. In Gosling et al's comprehensive and well-cited study~\cite{gosling2004should}, internet-submitted samples were found to be diverse, generalise across presentation formats, were not adversely affected by non-serious or repeat respondents, and present results consistent with findings from in-lab methods. There is similar evidence for usability experiments, in which  both  the  lab and  remote  tests captured similar information about the usability of websites~\cite{tullis2002empirical}.

That said, differences in results for lab and remote experiments are common~\cite{stern1997lost,buchanan2000potential,senior1999investigation}. The above website usability study also found that in-lab and remote experiments offered their own advantages and disadvantages in terms of the usability issues uncovered~\cite{tullis2002empirical}. The factors that influence differences between in-lab and remote research are still being understood, but even beyond experiment design, there is evidence that even aspects such as the participant-perceived geographical distance between the participant and the data collection system influences outcomes~\cite{moon1998effects}.

Reips'~\cite{reips2000web} well-cited study outlined 18 advantages of remote experiments, including (l) easy access to a demographically and culturally diverse participant  population, including participants from unique and previously inaccessible target populations; (2) bringing the experiment  to the participant instead of the opposite; (3) high statistical power by enabling access to large samples; (4) the direct assessment of motivational confounding; and (5) cost  savings of lab space, person-hours, equipment, and  administration. He found seven disadvantages: (l) potential for multiple submissions, (2) lack of experimental control, (3) participant self-selection, (4) dropout, (5) technical variances, (6) limited interaction with participants and (7) technical limitations.

\subsubsection{Supervised vs unsupervised}

With the increasing availability of teleconferencing, it has become possible for researchers to be co-"tele"present and supervise remote experiments through scheduling webcam experiment sessions. This presents a distinction from the unsupervised internet studies discussed above, and brings its own opportunities and limitations.  

Literature broadly suggests that unsupervised experiments provide suitable quality data collection~\cite{RYAN20131295,Hertzum2015WhatDT,Kettunen2018EffectsOU}. A direct comparison between a supervised in-lab experiment and a large, unsupervised web-based experiment found that the benefits outweighed its potential costs~\cite{RYAN20131295}; while another found that a higher percentage of high-relevance responses came from unsupervised participants than supervised ones in a qualitative feedback setting~\cite{Hertzum2015WhatDT}. There is also evidence that unsupervised participants react faster to tasks over the internet than those observed in the laboratory~\cite{Kettunen2018EffectsOU}.

For longitudinal studies, research in healthcare has found no significant difference between task adherence rates between unsupervised and supervised groups~\cite{creasey}. However, one study noted that supervised studies had more effective outcomes~\cite{Lacroix2015EffectsOA}.

\subsubsection{Crowdworkers: Viable?}

Remote data collection was theorised to bring easy access to participants, including diverse participants and large samples~\cite{reips2000web}. Researchers have found that recruiting crowdworkers, people who work on tasks distributed to them over the internet, allowed them access to a large participant pool\cite{paolacci2010running}, with enough diversity to facilitate cross-cultural and international research~\cite{buhrmester2016amazon}. Research has found that crowdworkers were significantly more diverse than typical American college samples and more diverse than other internet recruitment methods~\cite{buhrmester2016amazon}, at an affordable rate~\cite{paolacci2010running}\cite{buhrmester2016amazon}. This has allowed researchers a faster theory-to-experiment cycle~\cite{mason2012conducting}. 

Results from crowdworker-informed studies have been shown to reproduce existing results from historical in-lab studies ~\cite{paolacci2010running}~\cite{buhrmester2016amazon}~\cite{sprouse2011validation}, while a direct comparison between experiment groups of crowdworkers, social media-recruited participants and on-campus recruitment, found almost indistinguishable results~\cite{casler2013separate}. 

Some distinctions between crowdworkers and in-lab have been discovered, however. Comparative experiments between crowdworkers and in-person studies have suggested slightly higher participant rejection rates~\cite{sprouse2011validation}, while participants have been shown to report shorter narratives than other groups of college students (both online and in-person) and use proportionally more negative emotion terms than college students reporting verbally to an experimenter~\cite{grysman2015collecting}. 

Distinctions also exist within crowdworker recruitment sources. A study of AMT, CrowdFlower (CF) and Prolific Academic (ProA) found differences in response rate, attention-check question results, data quality, honesty, diversity and how successfully effects were reproduced~\cite{peer2017beyond}.

Data quality is a common concern regarding crowdworkers~\cite{goodman2013data}. However, attention-check questions used to screen out inattentive respondents or to increase the attention of respondents have been shown to be effective in increasing the quality of data collected~\cite{aust2013seriousness}, as have participant reputation scores~\cite{peer2014reputation}. 

A growing concern regarding crowdworkers is non-naivete, in which participants having some previous knowledge of the study or similar studies that might bias them in the experiment. Many workers report having taken part in common research paradigms~\cite{paolacci2014inside}, and there are concerns that if researchers continue to depend on this resource, the problem may expand. As such, further efforts are needed by researchers to identify and prevent non-naive participants from participating in their studies~\cite{buhrmester2018evaluation}.

\subsection{Summary}
It is clear that remote methods have been usefully deployed for non-XR research, and seemingly bring benefits such as easier participant recruitment, reduced recruitment cost and broadened diversity, without introducing major biases. However, there is still a paucity of research regarding the extent to which remote XR research can and has been used to leverage the unique benefits of both XR (environmental control, sensory illusions, data collection, replication) and remote (participation, practicality, cost-savings) methods, as well as the potential impact of their combined limitations. Therefore a survey of XR researcher experiences and beliefs regarding remote XR research could help us understand how these apply practically at the current time, and understand the key areas for future developments in this field. 


\section{Methodology}

\subsection{Survey}

We surveyed current practice to outline the researcher-perceived benefits and drawbacks of lab-based and remote XR research. We used a 30-item qualitative questionnaire that enquired about participants' existing lab-based and remote research practices; thoughts on future lab-based and remote research; and potential benefits and drawbacks for each area. The survey was circulated through relevant mailing lists (visionlist@visionscience.com, BCS-HCI@ jiscmail.ac.uk, chi-announcements@listserv.acm.org), to members of groups thinking of or currently running remote studies, and to members of universities' virtual and augmented reality groups found via search engines.

Responses were thematically analysed using an inductive approach based upon Braun and Clarke's six phases of analysis \cite{braun2006using}. The coding and theme generation process was conducted twice by independent researchers; themes were then reviewed collaboratively to create the final categorisations.

\subsection{Participants}

We received 46 responses to our survey from 36 different (predominantly academic) institutions. Most responses came from researchers based in Europe and North America, but responses also came from Asia. The majority of participants were either PhD students (18) or lecturers, readers or professors (11) at universities. Other roles were academic/scientific researcher (5), masters student (5), corporate researcher (4) and undergraduate student (2).  A diverse set of ages responded to the survey: 18-24 (5), 25-34 (22), 35-44 (11), 45+ (6), and gender skewed male (29) over female (16) or other (1).

\section{Participant XR setup results}

\begin{figure}[b]
\centering
\includegraphics[width=0.5\columnwidth]{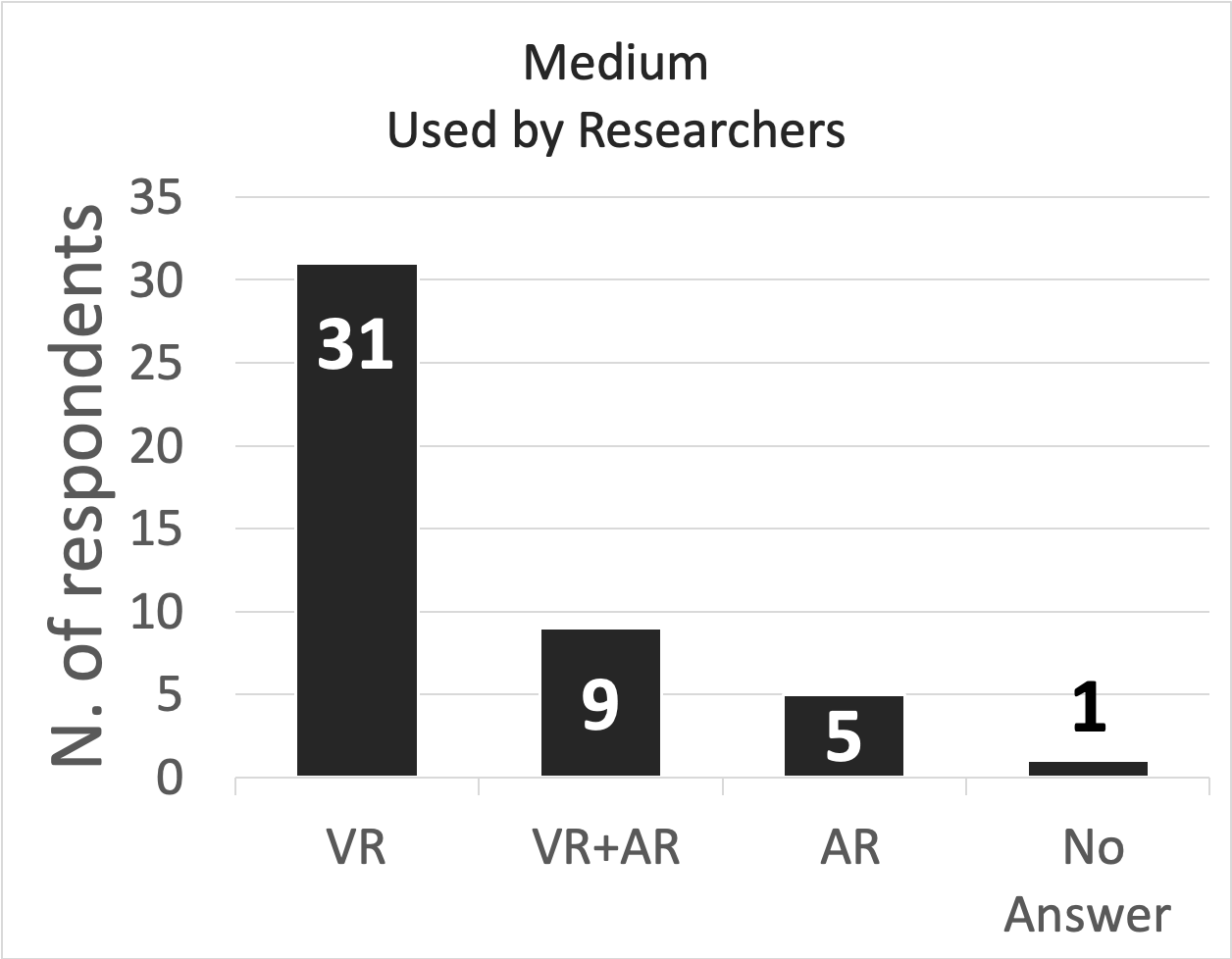}
\caption{Type of XR medium explored by survey respondents.}
\label{fig1}
\end{figure}

\begin{figure}[b]
\centering
\includegraphics[width=0.9\columnwidth]{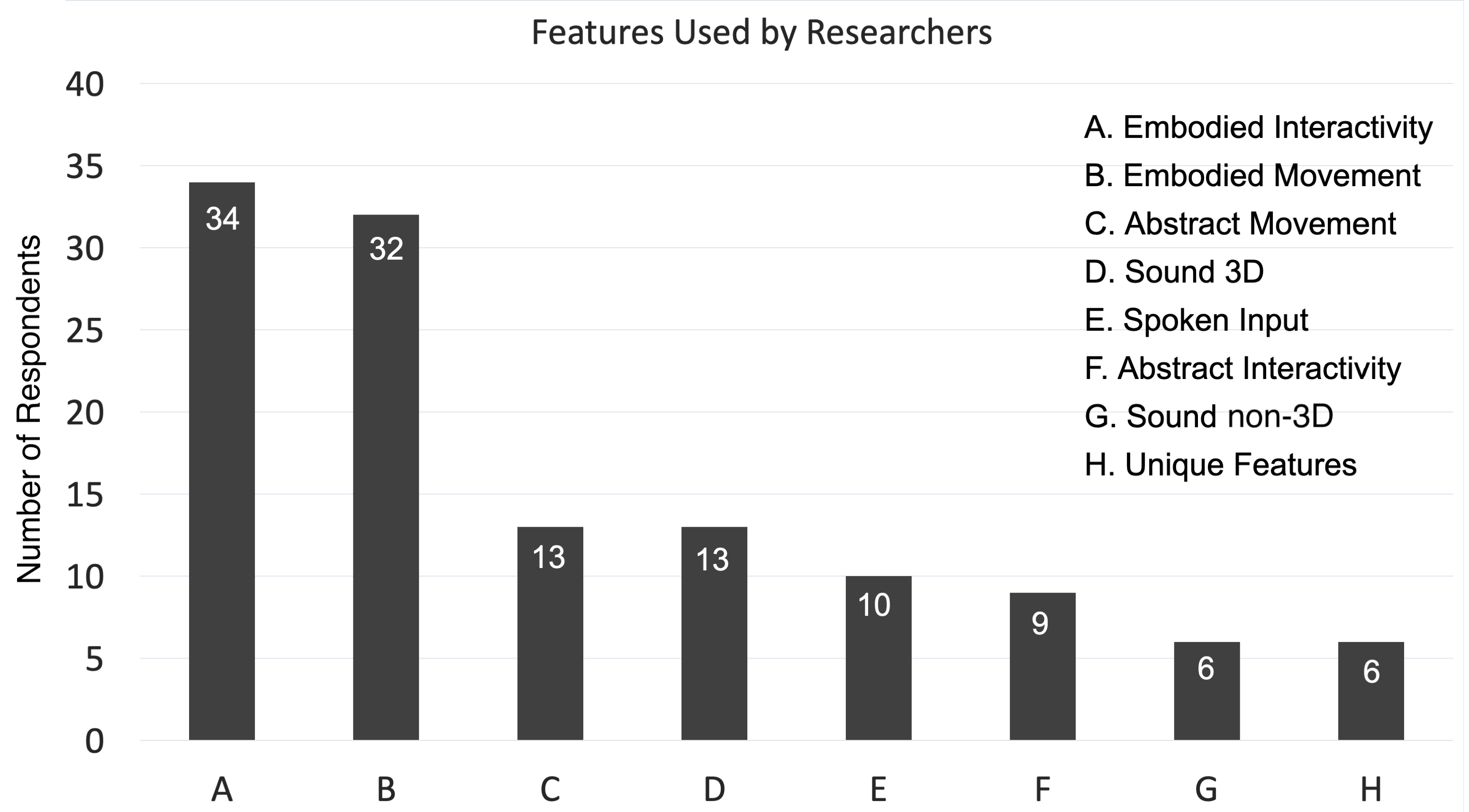}
\caption{Features used by respondents in their user studies. (A) Embodied Interactivity: using embodied controller/camera-based movement. (B) Embodied Movement:using your body to move/"roomscale". (C) Abstract Movement: using a gamepad or keyboard and mouse to move. (D) Sound 3D: binaural acoustics. E) Spoken Input. (F) Abstract Interactivity: using a gamepad or keyboard and mouse to interact. (G) Sound non-3D: mono/stereo audio. (H) Unique features: e.g. haptics, hand tracking, scent.}
\label{fig2}
\end{figure}

Participants were more likely to have previously ran in-lab studies (37) than remote studies (14). Twenty-seven participants noted that, because of the Covid-19 pandemic, they have considered conducting remote XR experiments. In the next six months, more researchers were planning to run remote studies (24) than lab-based (22).

Participants predominantly categorised their research as VR-only (28) over AR-only (5). Ten participants considered their research as both VR and AR (and three did not provide an answer). This result is illustrated in Fig.~\ref{fig1}. In terms of research hardware, the majority of VR research leveraged virtual reality HMD-based systems with six degrees of freedom (32), that tracks participants' movements inside the room, over three degrees of freedom (15) or CAVE systems (1). Nineteen researchers made use of embodied or gesture controllers, where the position of handheld controllers are tracked in the real world and their position virtualised. For AR, HMDs were the predominant medium (13) over smartphones (9), with some researchers (5) using both.

An array of supplementary technologies and sensors were also reported by 13 respondents, including gaming joysticks, haptic actuators, a custom haptic glove, motion capture systems, e-textiles, eye-trackers, microphones, computer screens, Vive body trackers, brain-computer interfaces, EEG and electrocardiogram (ECG) devices, galvanic skin response sensors and hand-tracking cameras, as well as other spatial audio and hardware rigs.

The use of a variety of different off-the-shelf systems was also reported: Vive, Vive Pro, Vive Eye, Vive Index,  Vive Pucks, Quest, Go, Rift, Rift S, DK2, Cardboard, Magic Leap One, Valve Knuckles, Hololens. Predominantly used devices are part of HTC Vive (25) and Oculus (23) family.

Respondents outlined numerous features of immersive hardware that they used in their research, visible in Fig. \ref{fig2}. The most prominent were embodiment aspects, including embodiment interactivity, in which a user's hand or body movements are reflected by a digital avatar (37) and embodiment movement (35), where participants can move in real space and that is recognised by the environment. Abstract movement (13), where a user controls an avatar via an abstracted interface (like a joystick) and abstract interactivity (8) were less popular. Spoken input was also used (10), as well as 3D sound (13) and non-3D sound (6). Scent was also noted (1) along with other unique features.

\section{Thematic analysis results}

In this section, we present and discuss the themes found in our survey study. The key points of each theme are summarised in a table at the start each subsection. Some of these points were found across multiple themes as they touch various aspects of user-based XR research.

\subsection{Theme: Study Sub-types}

\begin{table}[t]
  \caption{Summary of XR Study Sub-types}
  \label{tab:type}
  \begin{tabular}{p{2.5cm}p{5.5cm}}
    \toprule
    Method & Summary\\
    \midrule
    In-lab (vital) & Experiment requires features only feasible in-lab, e.g. bespoke hardware, unique data collection\\
    \\ [-3.5mm] 
    In-lab (preferred) & Concerns about integrity of data collected remotely, high value on controlled setting\\
    \\ [-3.5mm]
   
    Remote (vital) & User's natural (in-the-wild) environment is important (e.g. Social VR, naturally experienced at home and online)\\
    \\ [-3.5mm]
    
    Remote (preferred) & Priority to get cross-cultural feedback or reach large number of participant; lab provides limited benefits\\
  \bottomrule
\end{tabular}
\end{table}

Our analysis suggests that in-lab and remote studies can be additionally distinguished by whether the setting type is vital or preferred (summarised in Table \ref{tab:type}). Broadly, \textit{in-lab (vital)} studies require experimental aspects only feasible in-lab, such as bespoke hardware or unique data collection processes; \textit{in-lab (preferred)} studies could take place outside of labs, but prefer the lab-setting based upon heightened concerns regarding the integrity of data collected and place a high value on a controlled setting. \textit{Remote (vital)} studies are required when a user's natural environment is prioritised, such as explorations into behaviour in Social VR software; and \textit{remote (preferred)} studies are used when cross-cultural feedback or a large number of participants are needed, or if the benefits offered by an in-lab setting are not required.

Beyond these, another sub-type emerged as an important consideration for user studies: \textit{supervised} or \textit{unsupervised}. While less of an important distinction for in-lab studies (which are almost entirely supervised), participant responses considered both \textit{unsupervised} "encapsulated" studies, in which explanations, data collection and the study itself exist within the software or download process, and \textit{supervised} studies, in which researchers schedule time with the remote participant to organise, run and/or monitor the study. These distinctions will be discussed in more detail throughout the analysis below, as the sub-types have a distinct impact on many of the feasibility issues relating to remote studies.

\subsection{Theme: Study Participants}
\begin{table*}[t]
  \caption{Study Participants Key Points}
  \label{tab:users}
  \begin{tabular}{p{2.5cm}p{2cm}p{5cm}p{6.5cm}}
    \toprule
    Key Point&Issue&Lab&Remote\\
    \midrule
    Recruitment Scope&Sample size&Usually smaller numbers&Potential for larger number\\
    \\ [-3.5mm]
    Recruitment Scope&Sample balance&Might be easier to ensure balance&How to ensure balance? (e.g. who mostly owns XR equipment?)\\
    \\ [-3.5mm]
    Efficiency&Time&Requires setup time and organise participants&Potential less time especially if encapsulated and unsupervised\\
    \\ [-3.5mm]
    Precursor Requirements&Requisites&Pre-test and linguistic/culture comprehension conditions are ensured&Not clear how to verify conditions in remote studies\\
  \bottomrule
\end{tabular}
\end{table*}
\subsubsection{Recruitment scope}

Twenty-nine respondents stated the well-known challenge of recruiting a satisfactory number of participants for lab-based studies. Issues were reported both with the scale of available participants, and the problem of convenience sampling and WEIRD - Western, educated, industrialized, rich and democratic societies -  participants\cite{henrich2010most}. 

Participant recruitment was mentioned by 27 respondents as the area in which remote user studies could prove advantageous over labs. Remote studies could potentially provide easier recruitment (in terms of user friction: accessing to lab, arriving at the correct time), as well as removing geographic restrictions to the participant pool.

Removing the geographic restrictions also simplifies researchers' access to cross-cultural investigations (R23, R43). While cross-cul-\\tural lab-based research would require well-developed local recruitment networks, or partnerships with labs in target locations, remote user studies, and more specifically, systems built deliberately for remote studies, introduce cross-cultural scope at no additional overhead.

There are, however, common concerns over the limitations to these benefits due to the relatively small market size of XR technologies. For AR, this is not a strong  limitation for smartphones-based explorations, but the penetration of HMD AR and VR technology is currently limited, and it is possible that those who currently have access to these technologies will not be representative of the wider populations. Questions remain over who the AR/VR HMD owners are, if they exhibit notable differences from the general population, and if those differences are more impactful than those presented by existing convenience sampling.

Despite the belief that designing for remote participants will increase participant numbers, and therefore the power of studies, it seems unclear how researchers will reach HMD-owning audiences. Thirty respondents who have, or plan to, run remote XR studies have concerns about the infrastructure for recruiting participants remotely. Unlike other remote studies, the requirement for participants to own or have access to XR hardware greatly reduces the pool (around 5 million XR HMDs were sold in 2020 \cite{tankovska_2020}). A major outstanding question is how researchers can access these potential participants, although some platforms for recruiting XR participants have emerged in the past few months such as XRDRN.org.

Nine respondents noted that remote XR experiments may encourage participation from previously under-represented groups, including introverts and those who cannot or do not wish to travel into labs to take part (e.g. people who struggle to leave their homes due to physical or mental health issues).

However, respondents with research-specific requirements also raised concerns that recruitment of specific subsets of participants could be more difficult remotely. For example, when recruiting for a medical study of those with age-related mobility issues, it is unlikely that there will be a large cohort with their own XR hardware.

\subsubsection{Theme: Efficiency}

Twenty-five respondents noted the potential for remote studies to take up less time, particularly if remote studies are encapsulated and unsupervised. They stated that this removes scheduling concerns for both the researcher and the participant, and allows experiments to occur concurrently, reducing the total researcher time needed or increasing the scale of experiment. However, there are concerns this benefit could be offset by increased dropouts for longitudinal studies, due to a less "close" relationship between research and participant (R17, R25). 

\subsubsection{Participant precursor requirements}

One respondent noted they needed to run physiological precursor tests (i.e. visual acuity and stereo vision) that have no remote equivalent. Transitioning to remote research has meant this criteria must now be self-reported. Similarly, experiments have general expectations of linguistic and cultural comprehension, and opening research to a global scale might introduce distinctions from typically explored population. One respondent cautioned that further steps should be taken to ensure participants are able to engage at the intended level, as in-lab these could be filtered out by researcher intuition.

\subsection{Data Collection}

\begin{table*}[t]
  \caption{Data Collection Key Points}
  \label{tab:data}
  \begin{tabular}{p{2cm}p{7cm}p{7cm}}
    \toprule
    Key Point&Lab&Remote\\
    \midrule
    Hardware&Access custom and/or reliable hardware&Limited access to devices (e.g. EEG, ECG, computational power, etc.)\\
    \\ [-3.5mm]
    Data&Collection can be supervised, more detailed, real-time, more space for qualitative&Mostly unsupervised (less control), human expressions (e.g. facial) are generally lost, qualitative feedback is harder to collect\\
    \\ [-3.5mm]
    Behaviour&Likely more serious, richer (qualitative) data&Lack of detailed feedback, potentially less honest\\
  \bottomrule
\end{tabular}
\end{table*}

The overwhelming drawback of remote XR research, as reported by the majority respondents, was that of data collection. Excluding changes to participant recruitment, as mentioned above, the issues can broadly be categorised as: (1) bespoke hardware challenges, (2) monitoring/sensing challenges, and (3) data transmission and storage.

The use of bespoke hardware in any type of remote user study is a well-known issue, predominantly regarding the difficulty of managing and shipping bespoke technology to participants and ensuring it works in their test environments. In the context of XR technologies, 13 respondents voiced concerns about the complicated and temperamental system issues that could arise, particularly surrounding the already strenuous demands of PC-based VR on consumer-level XR hardware, without additional overheads (e.g. recording multiple cameras).

Four respondents felt it was unreasonable to ask remote participants to prepare multiple data-collection methods that may be typical in lab-studies, such as video recording and motion tracking. There were also concerns regarding the loss of informal, ad-hoc data collection (e.g. facial expressions, body language, casual conversations).

Finally, concerns were also raised regarding the efforts required to encapsulate all data capture into the XR experience, the effects this might have on data collection (for example, a recent study highlighted a difference on the variability of presence when participants recorded it from inside the VR experience versus outside~\cite{schwind2019using}), the reliability of transferring large amounts of data from participants, and how sensitive information (especially in the context of medical XR interventions) can securely be transferred and stored. This areas perhaps presents the biggest area for innovation for remote XR research, as it is reasonable to assume the academic community could create efficient, easy-to-use toolkits for remote data collection in XR environments which integrate to ethics-compliant data archives. 

Many data collection methods were deemed infeasible for remote experimentation: EEG, ECG, eye/hand tracking, GSR, as well as body language and facial expressions. Five researchers noted adaptions they had been working on to overcome these, including using HMD orientation to replace eye tracking, and using built-in HMD microphones to record breaths instead of ECG monitoring to determine exertion, or using the HMD controllers to perform hand tracking.

Respondents also noted some behavioural concerns and changes for remote, unsupervised participants. These included a lack of participation in qualitative feedback (6 respondents); for one researcher (R20), participants were "encouraged to provide feedback but few took the initiative." Another researcher (R31) stated "Debriefing is such a good space to collect unstructured interview data. Users relax after the questionnaire/debriefing ... produc[ing] a ... meta-narrative where participants consider your questions and their experiences together". The lack of supervision raised concerns regarding whether participants were being "truthful" in their responses, with one researcher (R41) stating that participants attempted to "game" their study in order to claim the participation compensation. However, others stated that unsupervised studies could reduce research bias arising from their perception of the participants' appearance and mannerisms. 

\subsection{Theme: Experiment Processes}

\begin{table*}[t]
  \caption{Experiment Process Key Points}
  \label{tab:exp}
  \begin{tabular}{p{3cm}p{1.5cm}p{5.5cm}p{5.5cm}}
    \toprule
    Key Point&Issue&Lab&Remote\\
    \midrule
    Process \& Guidance&Control&Full control over setup and participants&No control and guidance over participants\\
    \\ [-3.5mm]
    Process \& Guidance&Participants&Rapport with researcher, welcoming, more serious, attentive&Different attitude, potential cheating\\
    \\ [-3.5mm]
    Environment&Setting&Can be distracting (e.g. outside noise) but generally more controlled&Might be distracting or overwhelming but likely more realistic/natural for participants\\
    \\ [-3.5mm]
    Hardware \& software&Hardware&Access to custom devices, normal calibration process&No calibration (by researcher), potential for unknown errors, no custom tools\\
    \\ [-3.5mm]
    Hardware \& software&Software&Allows for Wizard of Oz, adjust setting in real time&Issues harder to spot and influence results, longer development time\\
    \\ [-3.5mm]
    Research questions&Topics&Unchanged, if we go back to normal research conditions&Remote setup might influence research questions and topics\\
    \\ [-3.5mm]
    Cost&Expenditures&More time consuming, more expensive to run&Potentially cheaper but potentially more work for implementation\\
  \bottomrule
\end{tabular}
\end{table*}

\subsubsection{Process \& Guidance}
Many respondents were concerned that unsupervised participants may conduct the experiments incorrectly, or have incorrect assumptions, or misunderstand processes or target actions. Twenty-four respondents felt that guidance would be better provided (introduction, explanations, etc) in a lab setting that also allows ad-hoc guidance and real-time corrections. 

There were also concerns over the mental state of participants: remote participants "may not take it seriously" or not focus (lack of motivation and engagement) or approach the study with a specific mood unknown to the researcher (R19, R30). Contrasting opinions suggested that participants may feel that the in-lab experience is "overly formal and uncomfortable" (R32).

Some respondents stated that remote experiments risk losing the "rapport" between researcher and participant, which might negatively influence the way a participant performs a remote study. However, one respondent stated that the transition to remote experimentation allowed them different, deeper, on-going connection with their participants. Their research was for a VR machine learning tool, and they found that moving away from in-person experimentation and to a remote workshop process encouraged the up-take of longitudinal community-building tools. The chosen communication method between researcher and user - Discord servers - became a place for unsupervised interaction between participants, and led to an on-going engagement with the research (R33). However it should be considered that any "rapport" between participant and researcher might introduce bias. 

\subsubsection{Environment}
Concern was raised around participants' environments, and their potential varying unsuitability for remote experimentation, compared with controlled laboratory settings. For example, one respondent (R20) stated: "one user reported walking into their whiteboard multiple times, causing low presence scores." The concern is particularly strong for unsupervised remote experiments, as distractions could enter into the experiment and affect data without the researcher being aware. 

This concern was not universal, however. Four respondents noted that their laboratories space was far from distraction free, and even suggested that a remote space could prove freer of interruptions than the space available to them in their research setting; while others stated that researchers should be mindful that the laboratory itself is an artificial space, far more so than where people will typically use their VR setups - in their homes. Five respondents highlighted how XR research could benefit from being deployed in "the participants' own environment".

The immediate environment of the user was also raised as a concern for VR experiment design: the choice of being able to move freely in an open space in a laboratory against a more adaptive solution for the unknown variables of participants' home environments.

Respondents noted that supporting the different VR and AR setups to access a larger remote audience would also prove more labour-intensive, and would introduce more variables compared with the continuity of the tech stack available in-lab. With remote experiments, and more so for encapsulated unsupervised ones, 10 respondents believe there will be more time spent in developing the system.

\subsubsection{Hardware and software}

A concern regarding remote experiments, particularly unsupervised, is that calibration processes are harder to verify (R30). This could either cause participants to unknowingly have faulty experiences, and therefore report faulty data; or it will increase time taken to verify user experiences are correct. Unknown errors can effect data integrity or participant behaviour. Respondents noted that this type of remote error are often much more difficult and labour-intensive to fix compared with in-lab. This issue is compounded by individual computer systems introducing other confounding factors (for both bug-fixing and data collection) such as frame-rates, graphic fidelity, tracking quality and even resolution can vary dramatically. 

Five respondents reflected that overcoming these issues could lead to more robust research plans, as well as better development and end-product software to overcome problems listed. This encapsulation could also lead to easier opportunities for reproducability, as well as the ability for researchers to share working versions of the experiment with other researchers, instead of just the results. It could also help with the versioning of experiments, allowing researchers to build new research on-top of previous experiment software.

Four respondents were aware these advantages are coupled with longer development times. The increased remote development requirements could also be limiting for researchers who face constrained development resources, particularly those outside of computer science departments. This is compounded by the fact that the infrastructure for recruiting remote XR participants, data capture, data storage and bug fixing is not particularly developed. Once these are established, however, respondents felt these might make for a higher overall data quality compared with the current laboratory-based status quo, due to more time spent creating automated recording processes, and not relying on researcher judgement. There are also arguments that the additional development time is offset by the potential increase in participants and, if unsupervised, the reduction in experiment supervision requirements.

Six respondents that use specific hardware in their research, noted that it was currently difficult to measure physiological information in a reliable way, and included hand tracking in this. However, we are aware that some consumer VR hardware (Oculus Quest) allows hand-tracking, and so there is an additional question of whether researchers are being fully supported in knowing what technologies are available to them.

To alleviate issues with reaching participants, two respondents wrote about potentially sending equipment to participants. The limitations of this were noted as hardware having gone missing (which had happened, R35), and participants being unable to use equipment on their own (which had not happened yet).

\subsubsection{Research questions}
Five respondents noted that their research questions changed or could change depending on whether they were aiming for a laboratory or remote settings. For example, one respondent (R31) suggested that "instead of the relationship of the physical body to virtual space, I'd just assess the actions in virtual space". Others explored the potentiality of having access to many different system setups, for example, now being able to easily ask questions like "are there any systematic differences in cybersickness incidence across different HMDs?". (R39)

Nine respondents speculated that remote research has potential for increasing longitudinal engagement, due to lower barriers to entry for researcher (room booking, time) and participant (no commute), and that rare or geographically based phenomena could be cheaply studied using remote research; as providing those communities access to VR may be cheaper than relocating a researcher to them.

\subsubsection{Costs}
Eight respondents noted the potential of remote experimentation for reducing some of the cost overheads for running experiments. Laboratories have important costs that are higher than remote studies: lab maintenance, hardware maintenance, staff maintenance. Without these, costs per participant are lower (and for unsupervised studies, almost nil). As experiment space availability was also noted as a concern for laboratory-based experiments, this seems a potentially under-explored area of benefit, provided remote participant recruitment is adequate.

\subsection{Theme: Health \& Safety}

\begin{table}[t]
  \caption{Health and Safety Key Points}
  \label{tab:hns}
  \begin{tabular}{p{2cm}p{5.5cm}}
    \toprule
    Key Point&Summary\\
    \midrule
    Protocols&Missing standard protocols (to work safely with participants in-lab)\\
    \\ [-3.5mm]
    Equipment&Sanitizing of in-lab equipment and spaces\\
    \\ [-3.5mm]
    Remote&Concerns for remote participants (e.g. accidents during a user study)\\
    \\ [-3.5mm]
    Real-Time Aid&Not available for remote participants (e.g. motion sickness)\\
  \bottomrule
\end{tabular}
\end{table}

The leading benefit given for remote user studies was that of health and safety, citing shared HMDs and controllers as a potential vector for Covid-19 transmission, as well as more general issues such as air quality in enclosed lab spaces. Concerns were raised for both viral transmission between participants, and between participant and the researcher. This concern has also increased administration overheads, with 6 respondents stating it could be more time consuming to prepare the lab and organise the studies or using new contract-tracing methods for lab users.

However, respondents also raised concerns about additional safety implications for remote participants. The controlled lab environment is setup to run the study, whereas remote participants are using a general-purpose space. One AR researcher who conducts research that requires participants to move quickly outside in fields noted his study could be considered "incredibly unsafe" if unsupervised or run in an inappropriate location. Additionally, for health and mental health studies, in-lab allows for researcher to provide support, especially with distressing materials. Finally, VR environment design has a direct impact on the level of simulator sickness invoked in participants. There were questions about the responsibility of researchers to be present to aid participants who could be made to feel unwell from a system they build.

\subsection{Theme: Ethics}

Three ethics concerns were reported by respondents: encouraging risky behaviours, responsibility for actions in XR and data privacy. An example of this might be the ethical implications of paying participants, and therefore incentivising them, to take part in what could be considered a high-risk behaviour: entering an enclosed space with a stranger and wearing a VR HMD.

Respondent (R30) raised the question of liability for participants who are injured in their homes while taking part in an XR research project. The embodied nature of XR interventions - and most respondents used this embodiment in their studies - could put participants at a greater risk of harming themselves than with other mediums.

Finally, while cross-cultural recruitment was seen as a potential boom for remote research, questions were raised about ethics and data storage and protection rules when participants are distributed across different countries, each with different data storage laws and guidelines. Although not limited to XR, due to the limited number of VR users, and the disproportionate distribution of their sales, it seems the majority of remote VR participants will originate from North America, and ethics clarification from non-US-based universities are needed.

\section{Covid-19 Implications}

\begin{table}[t]
  \caption{Covid-19 Implications}
  \label{tab:cov}
  \begin{tabular}{p{2cm}p{5.5cm}}
    \toprule
    Key Point&Summary\\
    \midrule
    Suspensions&No user studies at the moment\\
    \\ [-3.5mm]
    Facilities & Sanitizing of equipment and spaces\\
    \\ [-3.5mm]
    Recruitment&Harder/impossible to recruit in-lab participants\\
    \\ [-3.5mm]
    Exclusion&Bias and high risk participants\\
  \bottomrule
\end{tabular}
\end{table}

While Covid-19 has impacted most studies around the world, the dependence on shared hardware for XR research, especially HMDs, has led to many implications reported by our respondents. These concerns are particularly related to Covid-19, and therefore be reduced as the pandemic is resolved. However, as it is currently unclear when the pandemic will end, we felt it was useful to discuss them in a dedicated section.

Most respondents noted that Covid-19 had caused a suspension of studies and that they were unclear how long the suspension would last for, resulting in an overall drop in the number of studies being conducted, with 30 respondents stating it will change the research they conduct (e.g. moving to online surveys). The continuation of lab studies was eventually expected, but with added sanitizing steps. However for many, it was unclear what steps they should take in order to make XR equipment sharing safe. These concerns extended beyond the XR hardware to general facility suitability, including room airflow and official protocols which may vary for each country and/or institution.

Five respondents also had concerns about participants. There were worries that lab-based recruitment would be slow to recover, as participants may be put off taking part in experiments because of the potential virtual transfer vectors. Similarly, respondents were concerned about being responsible for participants, and putting them in a position in which their is a chance they could be exposed to the virus.

There was also concerns around Covid-19 and exclusion, as researchers who are at high risk of Covid-19 or those who are in close contact with high risk populations, would now have to self-exclude from lab-based studies. This might introduce a participant selection bias towards those willing to attend a small room and sharing equipment,

It should be noted that not all labs are facing the same problems - some of our respondents had continued lab-based experimentation during this period, with Covid-19 measures ensuring that participants wore face masks during studies. This was considered a drawback as combined with an HMD, it covered the participant's entire face and was cumbersome. These measures are also known not to be 100\% protective.

\section{Discussion}
In the previous section we presented the results as themes we found in our analysis. Some of these presented common characteristics and some issues were reported in multiple themes. We now summarise the results, highlight the key points and suggest important questions for future research.

\subsection{Recruitment and participants}
As with non-XR experiments, researchers are interested in the potential benefits of remote research for increasing the amount, diversity and segmentation of participants compared with in-lab studies. However, with many respondents reporting that it has been difficult to recruit XR participants, it seems there is a gap between potential and practice. The unanswered question is how to build a pool of participants that is large and diverse enough to accommodate various XR research questions and topics, given that there are few high-end HMDs circulating in the crowdworker community \cite{ma2018web}\cite{mottelson2017virtual}. So far, we have found three potential solutions for participant recruitment, although each requires further study:

(1) Establish a dedicated XR crowdworker community. However, concerns of non-naivety\cite{paolacci2014inside}, which are already levied at the much larger non-XR crowdworker participant pools, would surely be increased. We would also have to understand if the early version of this community would be WEIRD\cite{henrich2010most} and non-representative, especially given the cost barrier to entry for HMDs.

(2) Leverage existing consumer XR communities on the internet, such as the large discussion forums on Reddit. These should increase in size further as they shift from early-adopter to general consumer communities. However, these communities may also have issues with representation.

(3) Establish hardware-lending schemes to enable access to a broader base of participants~\cite{steedartticle}. However, the cost of entry and risk of these schemes may make them untenable for smaller XR research communities.

It is also not clear, beyond HMD penetration, what the additional obstacles are that XR poses for online recruitment. Technical challenges (e.g. XR applications needing to run on various devices, on different computers, requiring additional setup beyond simple software installation) and unintuitive experiment procedures (e.g. download X, do an online survey at Y, run X, record Z) for participants are notable distinct issues for remote XR research. It is also unclear if the use of XR technology has an impact on what motivates participants to take part in remote studies, an area of study that has many theoretical approaches even in the non-XR area\cite{keusch2015people}.

\subsection{Data collection}
Respondents feel that many types of physiological data collection are not feasible with either XR or non-XR remote research. For remote XR research, there are unique concerns over video and qualitative data collection  as using XR technologies can make it (technically) difficult to reliably video or record the activity, as well as moving participants' loci of attention away from the camera or obscuring it behind an HMD. However, the hardware involved in creating XR experiences provides a variety of methods to gather data, such as body position, head nodding, breath-monitoring, hand tracking, HMD angle instead of eye tracking. These can be used to explore research topics that are often monitored via other types of physiological, video or qualitative data, such as attention, motivation, engagement, enjoyment, exertion or focus of attention. It would be useful for XR researchers to build an understanding of what the technologies that are built into XR hardware can tell us about participant experiences, so as to allow us to know the data collection affordances and opportunities of XR hardware.

That said, the infrastructure for collecting and storing this (mass) of XR data remotely is currently not fully implemented, and we are not aware of any end-to-end standardised framework. However, work is being done to simplify the data collection step for XR experiments build in Unity~\cite{brookes2019studying}. There are also opportunities to further develop web-based XR technologies that could send and store data on remote servers easily. There are also ethical concerns, as respondents were unclear on guidance regarding data collection from participants located in other nations, particularly when they should be paid. This includes how the data is collected, where it should be stored, and how can be manipulated.

\subsection{Health, safety and Covid-19}
At the time of writing, many laboratories are considered unsafe for running user studies. Although some respondents reported being able to work in-lab, the limitations mean it is not currently feasible to run user studies under normal conditions. The main concern for the near future is the lack of standardised protocols to ensure safety of researchers and participants while running user studies and the issue with the ethics protocols of the research institutions. For XR research, it is unclear how to adequately sanitize equipment and tools, as well as how to maintain physical distancing. There are also concerns about the comfort of participants if they are required to wear masks alongside HMDs. Finally, respondents reported concerns about a potential long-term fall in user motivation to take part in such experiments, when HMDs are a notable infection vector. There are distinctly different safety and ethics concerns around remote XR experiments, including the research responsibility for not harming participants (e.g. ensuring environments are safe for the movements, and not inducing simulator sickness), which, while also true of in-lab experiments, are considered a greater challenge when a participant is not co-located with the researcher.

\subsection{Mediated impact}

Respondents reported framing their research questions and experiments differently depending on the target experiment setting. The strongest transition was that of an in-lab study of participants using an AR HMD (Hololens), which changed to a remote study that had participants watch a pre-recorded video of someone using the AR HMD. It seems these kinds of transitions will continue to be necessary depending on how esoteric the hardware is, with fewer concerns for AR smartphone investigations. 

A concern for respondents was that remote settings introduce additional uncontrolled variables that need to be considered by researchers, such as potential unknown distractions, trust in participants and their motivation, and issues with remote environmental spaces. However, previous research shows that most HMD-wearing remote participants engage in space well-known to them (the home) and predominantly when they are alone~\cite{ma2018web}, which could alleviate some of the environmental space and distraction concerns. Further research into how a home environment could impact XR studies is needed, and the creation of well-defined protocols to alleviate uncontrolled influences remote XR results. Beyond this, we also need to understand any impact that remote experiments may have on results compared with in-lab experiences, especially if we are to be able to reliably contrast lab and remote research. Previous research for non-XR experiments suggest that distinctions between lab and remote settings exist~\cite{buchanan2000potential}\cite{senior1999investigation}~\cite{stern1997lost}, but it has been theorised that the impact might be less for XR experiments, as you "take the experimental environment with you"~\cite{blascovich2002immersive}.

\subsection{"Encapsulated" experiments: the ideal?}
Respondents stated that creating remote XR experiments might encourage better software development and experimental processes. If experiments are able to be deployed as all-in-one experience and data collection bundles that can run unsupervised, the time-saving implications for researchers (and participants) are huge, especially when paired with the potential increase in participants. This type of "encapsulated experiment" can also improve replication and transparency, as theorised by Blascovich~\cite{blascovich2002immersive}, and allow for versioning of experiments, in which researchers can build on perfect replicas of other's experimental environments and processes. Finally, due to the similar nature of XR hardware, data logging techniques could easily be shared between system designers or standardised; something we have seen with the creation of the Unity Experiment Framework~\cite{brookes2019studying}.

However, there are some limitations to this approach. It is likely it will require additional development time from the researchers, especially as a comprehensive experiment framework is established. In addition, there are data collection limitations for remote XR studies, as discussed in previous sections. It is also interesting to consider how encapsulation might work for AR investigations, as the environment will only partially be controlled by the designer.

We believe that the potential for remote XR experiments lies in understanding the data collection affordances of the hardware; collectively building frameworks to ease the collection of this data; and to design research questions that maximise their use; all inside encapsulated experiences. This might be a mindset shift for researchers, who according to our survey, are predominantly lab-orientated.

\section{Limitations}

Our goal with this research was to provide an overall insight into the XR researcher community. However, this approach means that insights from sub-communities may not have been found. For example, we had no responses from researchers involved in topics such as vulnerable populations. Further investigation into sub-communities is needed to uncover potential insights for those areas.

\section{Conclusion and Recommendations}

It is clear from our survey that respondents believe that remote XR research has the potential to be a useful research approach. However, it currently suffers from numerous limitations regarding data collection, system development and a lack of clarity around participant recruitment. Analysis of our survey results and literature around remote and remote XR research suggest that, to better understand the boundaries of remote XR experimentation, researchers need answers to the following questions: 

\begin{itemize}
    \item [(1)] Who are the potential remote XR participants, and are they representative? 
    \item [(2)] How can we access a large pool of remote XR participants? 
    \item [(3)] To what extent do remote XR studies affect results compared with in-lab?
     \item [(4)] What are the built-in XR data collection affordances of XR hardware, and what can they help us study?
     \item [(5)] How can we lower the barriers to creating encapsulated experiment software, to maximise the potential of remote XR research?
    
\end{itemize}
We believe there is an opportunity to reconceptualise approaches to XR and remote research. XR experiments, as it stands, are predominantly used to study a participant's experience with an XR system, in an artificial but controlled setting (laboratory) using external data collection methods (surveys, cameras, etc.). However, if we consider XR devices primarily as data-collection hardware with set properties, we can work backwards to understand what research questions are suitable with the existing data collection afforded by XR hardware. Additionally, we also believe that there is potential to reconceptualise, for suitable applications, the home as a natural research location and move away from the laboratory as the default location for user studies. This is a potentially unique opportunity for XR compared with non-XR studies as, for many investigations, the XR experiment takes the environment with it.

\begin{acks}
This work is supported by the EPSRC and AHRC Centre for Doctoral Training in Media and Arts Technology (EP/L01632X/1).
\end{acks}

\bibliographystyle{ACM-Reference-Format}
\bibliography{sample-base}

\appendix

\end{document}